\documentstyle[12pt]{article}
\input tcilatex
\QQQ{Language}{
American English
}

\begin{document}

\title{{\Large {\bf DEDUCTION OF THE QUANTUM NUMBERS OF LOW-LYING STATES OF
6-NUCLEON SYSTEMS BASED ON SYMMETRY}}}
\author{{\small C.G.Bao$^1$, Y.X.Liu$^2$} \\
{\small $^1$ Department of Physics,Zhongshan University,
Guangzhou,510275,P.R.China.}\\
{\small $^2$ Department of Physics, Beijing University, Beijing, 100871,
P.R.China}}
\date{}
\maketitle

\hspace{1.0in}

ABSTRACT: The inherent nodal structures of the wavefunctions of 6-nucleon
systems are investigated. A group of six low-lying states (including the
ground states) dominated by total orbital angular momentum L=0 components
are found, the quantum numbers of each of these states are deduced. In
particular, the spatial symmetries of these six states are found to be
mainly the \{4,2\} and \{2,2,2\}.

\hspace{1.0in}


\hspace{1.0in}

\hspace{1.0in}

As a few-body system the 6-body system has been scarcely investigated
theoretically due to the complexity arising from the 15 spatial degrees of
freedom. The existing related literatures concern mainly the ground states
and a few resonances [1-5]. The study of the character of the excited states
is very scarce. On the other hand, the particles of 6-body systems are
neither too few nor too many. The study of them is attractive because it may
lead to an understanding of the connection between the few-body theory and
the modal theories for nuclei. Before solving the 6-body Schr\"odinger
equation precisely, if we can have some qualitative understanding of the
spectrum, it would be very helpful. This understanding , together with the
results from calculations and experiments, will lead to a complete
comprehension of the physics underlying the spectrum. In [6] the qualitative
feature of 4-nucleon systems has been studied based on symmetry. In this
paper we shall generalize the idea of [6] to extract qualitative character
of the low-lying states of 6-nucleon systems.

There are two noticeable findings in [6]. (i) The ground state is dominated
by total orbital angular momentum L=0 component, while all the resonances
below the 2n+2p threshold are dominated by L=1 components, there is a very
large gap lying between them. Experimentally, this gap is about 20 MeV. This
fact implies that the collective rotation is difficult to be excited. (ii)
The internal wavefunctions (the wavefunction relative to a body-frame) of
all the states below the 2n+2p threshold do not contain nodal surfaces. This
fact implies that the excitation of internal oscillation takes a very large
energy. Therefore, ti would be reasonable to assume that the L=0 nodeless
component will be also important in the low-lying spectrum of the 6-nucleon
systems.

It was found in [7,8] that a specific kind of nodal surfaces may be imposed
on the wavefunctions by symmetry. Let $\Psi $ be an eigenstate. Let $A$
denotes a geometric configuration. In some cases $A$ may be invariant to
specific combined operations $O_i$ ( i=1 to m). For example, when $A$ is a
regular octahedron (OCTA) for a 6-body system, then $A$ is invariant to a
rotation about a 4-fold axis of the OCTA by 90$^{\circ }$ together with a
cyclic permutation of four particles. In this case we have

\[
\stackrel{\wedge }{O_i}\Psi (A)=\Psi (O_iA)=\Psi (A)\hspace{1.0in}(1) 
\]

Owing to the inherent transformation property of $\Psi $ (the property with
respect to rotation, inversion, and permutation), (1) always can be written
in a matrix form (as we shall see) and appears as a set of homogeneous
linear algebra equations. They impose a very strong constraint on $\Psi $ so
that $\Psi $ may be zero at $A$. This is the origin of this specific kind of
nodal surfaces, they are called the inherent nodal surfaces (INS).

The INS appear always at geometric configurations with certain geometric
symmetry. For a 6-body system the OCTA is the configuration with the
strongest geometric symmetry. Let us assume that the six particles form an
OCTA. Let k' be a 4-fold axis of the OCTA, and let the particles 1,2,3, and
4 form a square surrounding k'. Let $R_\delta ^{k^{\prime }}$ denote a
rotation about k' by the angle $\delta $ (in degree), let $p(1432)$ denotes
a cyclic permutation. Evidently, the OCTA\ is invariant to

\[
O_1=p(1432)R_{-90}^{k^{\prime }}\hspace{1.0in}(2) 
\]

Let $p_{ij}$ denotes an interchange of the locations of particles i and j, $%
P $ denotes a space inversion. The OCTA is also invariant to

\[
O_2=p_{13}p_{24}p_{56}P\hspace{1.0in}(3) 
\]

Let i' be an axis vertical to k' and parallel to an edge of the above
square; say, parallel to $\stackrel{\rightarrow }{r_{12}}$. Then the OCTA\
is also invariant to

\[
O_3=p_{14}p_{23}p_{56}R_{180}^{i^{\prime }}.\hspace{1.0in}(4) 
\]

Let $OO^{\prime }$ be a 3-fold axis of the OCTA, where $O$ denotes the
center of mass. Let particles 2,5, and 3 form a regular triangle surroundung
the $OO^{\prime }$; 1,4, and 6 form another triangle. Then the OCTA is also
invariant to

\[
O_3=p(253)p(146)R_{-120}^{oo^{\prime }}\hspace{1.0in}(5) 
\]

Besides, the OCTA is also invariant to some other operators, e.g., the $%
p(152)p(364)R_{-120}^{oo"}$ (where $OO"$ is another 3-fold axis). However,
since the rotations about two different 3-fold axes are equivalent, one can
prove that this additional operator does not introduce new constraints, and
the operators $O_1$ to $O_4$ are sufficient to specify the constraints
arising from symmetry.

Let an eigenstate of a 6-nucleon system with a given total angular momentum
J, parity $\Pi $, and total isospin T be written as 
\[
\Psi =\sum_{L,S}\Psi _{LS}\hspace{1.0in}(6) 
\]
where S is the total spin, 
\[
\Psi _{LS}=\sum_{\lambda i}F_{LSM}^{\lambda i}\chi _S^{\stackrel{\symbol{126}%
}{\lambda }i}\hspace{1.0in}(7) 
\]
Where $M$ is the Z-component of L, $F_{LSM}^{\lambda i}$ is a function of
the spatial coordinates, which is the i$^{th}$ basis function of the $%
\lambda -$representation of the S$_6$ permutation group. The $\chi _S^{%
\stackrel{\sim }{\lambda }i}$ is a basis function in the spin-isospin space
with a given S and T and belonging to the $\stackrel{\sim }{\lambda }-$%
representation, the conjugate of $\lambda .$ In (7) the allowed $\lambda $
are listed in Table 1, they depend on S and T [9].

\hspace{1.0in}

\begin{tabular}{|c|c|c|}
\hline
S & T & $\lambda $ \\ \hline
0 & 0 & \{1$^6$\}, \{2,2,1,1\},\{3,3\},\{4,1,1\} \\ \hline
1 & 0 & \{2,1$^4$\}, \{3,1$^3$\}, \{2,2,2\}, \{3,2,1\}, \{4,2\} \\ \hline
2 & 0 & \{2,2,1,1\}, \{3,2,1\} \\ \hline
3 & 0 & \{2,2,2\} \\ \hline
0 & 1 & \{2,1$^4$\}, \{3,1$^3$\}, \{2,2,2\}, \{3,2,1\}, \{4,2\} \\ \hline
1 & 1 & \{1$^6$\}, \{2,1$^4$\}, 2\{2,2,1,1\}, \{3,1$^3$\}, 2\{3,2,1\},
\{3,3\}, \{4,1,1\} \\ \hline
2 & 1 & \{2,1$^4$\}, \{2,2,1,1\}, \{3,1$^3$\}, \{2,2,2\}, \{3,2,1\} \\ \hline
3 & 1 & \{2,2,1,1\} \\ \hline
\end{tabular}

Tab.1, The allowed representation $\lambda $ in (7)

\hspace{1.0in}

From k' and i' defined before one can introduce a body frame i'-j'-k'. In
the body-frame the $F_{LSM}^{\lambda i}$ can be expanded 
\[
F_{LSM}^{\lambda i}(123456)=\sum_Q D_{QM}^L(-\gamma ,-\beta ,-\alpha
)F_{LSQ}^{\lambda i}(1^{\prime }2^{\prime }3^{\prime }4^{\prime }5^{\prime
}6^{\prime })\hspace{1.0in}(8) 
\]
Where $\alpha \beta \gamma $ are the Euler angles to specify the collective
rotation, $D_{QM}^L$ is the well known Wigner function, Q are the projection
of L\ along the k'-axis. The (123456) and (1'2'3'4'5'6') specifies that the
coordinates are relative to a fixed frame and the body-frame, respectively.

Since the $F_{LSQ}^{\lambda i}$ span a representation of the rotation group,
space inversion group, and permutation group, the invariance of the OCTA to
the operations $O_1$ to $O_4$ leads to four sets of equations. For example,
from 
\[
\hat O_1F_{LSQ}^{\lambda i}(A)=F_{LSQ}^{\lambda i}(O_1A)=F_{LSQ}^{\lambda
i}(A)\hspace{1.0in} (9) 
\]
where $F_{LSQ}^{\lambda i}(A)$ denotes that the coordinates in $%
F_{LSQ}^{\lambda i}$ are given at an OCTA, for all $Q$ with $|Q|\leq L$ we
have 
\[
\sum_{i^{\prime }}[g_{ii^{\prime }}^\lambda (p(1234))e^{-i\frac \pi
2Q}-\delta _{ii^{\prime }}]F_{LSQ}^{\lambda i^{\prime }}(A)=0\hspace{1.0in}%
(10) 
\]
where $g_{ii^{\prime }}^\lambda $ are the matrix elements belonging to the
representation $\lambda $, which are known from the textbooks of group
theory (e.g., refer to [10]). From $\hat O_2$ and $\hat O_4,$ we have 
\[
\sum_{i^{\prime }}[g_{ii^{\prime }}^\lambda (p_{13}p_{24}p_{56})\Pi -\delta
_{ii^{\prime }}]F_{LSQ}^{\lambda i^{\prime }}(A)=0\hspace{1.0in}(11) 
\]
and 
\[
\sum_{Q^{\prime }i^{\prime }}[(-1)^Lg_{ii^{\prime }}^\lambda
(p_{14}p_{23}p_{56})\delta _{\stackrel{-}{Q}Q^{\prime }}-\delta _{ii^{\prime
}}\delta _{QQ^{\prime }}]F_{LSQ^{\prime }}^{\lambda i^{\prime }}(A)=0\ %
\hspace{1.0in}(12) 
\]
where $\stackrel{-}{Q}=-Q.$ It is noted that 
\[
R_{-120}^{oo^{\prime }}=R_\theta ^{j^{\prime }}R_{-120}^{k^{\prime }}R_{%
\stackrel{-}{\theta }}^{j^{\prime }}\hspace{1.0in}(13) 
\]
where $\theta =\arccos (\sqrt{\frac 13})$. Thus from $\hat O_3$ we have 
\[
\sum_{Q^{\prime }i^{\prime }}[g_{ii^{\prime }}^\lambda
[p(235)p(164)]\sum_{Q^{\prime \prime }}D_{QQ"}^L(0,\theta ,0)e^{-i\frac{2\pi 
}3Q"}D_{Q^{\prime }Q"}^L(0,\theta ,0)-\delta _{ii^{\prime }}\delta
_{QQ^{\prime }}]F_{LSQ^{\prime }}^{\lambda i^{\prime }}(A)=0\hspace{1.0in}%
(14) 
\]
Eq.(10), (11), (12), and (14) are the equations that the $F_{LSQ}^{\lambda
i}(A)$ have to fulfilled. In some cases there is one or more than one
nonzero solution(s) (i.e., not all the $F_{LSQ}^{\lambda i}(A)$ are zero) to
all these equations . But in some other cases, there are no nonzero
solutions. In the latter case, the $\Psi _{LS}$ has to be zero at the OCTA
configurations disregarding their size and orientation. Accordingly, an INS
emerges and the OCTA is not accessible. Evidently, the above equations
depend on and only on L, $\Pi $, and $\lambda .$ Therefore the existence of
the INS does not at all depend on dynamics (e.g., not on the interaction,
mass, etc.).

Since the search of nonzero solutions of linear equations is trivial, we
shall neglect the details but give directly the results of the L=0
components in the second and fourth columns of Tab.2

\hspace{1.0in}

\begin{tabular}{|c|c|c||c|c|}
\hline
& 0$^{+}$ & 0$^{+}$ & 0$^{-}$ & 0$^{-}$ \\ \hline
$\lambda $ & OCTA & C-PENTA & OCTA & C-PENTA \\ \hline
\{6\} & 1 & 1 & 0 & 0 \\ \hline
\{5,1\} & 0 & 1 & 0 & 0 \\ \hline
\{4,2\} & 1 & 1 & 0 & 0 \\ \hline
\{3,3\} & 0 & 1 & 0 & 0 \\ \hline
\{2,2,2\} & 1 & 1 & 1 & 0 \\ \hline
\{2,2,1,1\} & 0 & 1 & 0 & 0 \\ \hline
\{2,1$^4$\} & 0 & 1 & 0 & 0 \\ \hline
\{1$^6$\} & 0 & 1 & 0 & 0 \\ \hline
\{3,2,1\} & 0 & 2 & 0 & 0 \\ \hline
\{4,1,1\} & 0 & 0 & 0 & 0 \\ \hline
\{3,1$^3$\} & 0 & 0 & 1 & 0 \\ \hline
\end{tabular}

Tab.2, The accessibility of the OCTA (regular octahedron) and the C-PENTA
(regular centered-pentagon) to the L$^\Pi =0^{+}$ and $0^{-}$ wavefunctions
with different spatial permutation symmetry $\lambda $. Where the figures in
the blocks are the numbers of independent nonzero solutions. The figure 0
implies that nonzero solutions do not exist.

\hspace{1.0in}

The INS existing at the OCTA may even extend beyond the OCTA. For example,
when the shape in Fig.1a is prolonged along k', then the shape is called a
prolonged-octahedron. This shape (denoted by $B$ ) is invariant to $O_1,O_2,$
and $O_4$, but not to $O_3$. Hence, the $F_{LSQ^{\prime }}^{\lambda
i^{\prime }}(B)$ should fulfill only (10) to (12), but not (14). When
nonzero common solutions of (10), (11), (12), and (14) do not exist, while
nonzero solutions of only (10) to (12) also do not exist, the INS extends
from the OCTA to the prolonged-octahedrons. An OCTA has many ways to
deform;e.g., instead of a square, the particles 1,2,3, and 4 form a
rectangle or form a diamond, etc.. Hence, the INS at the OCTA has many
possibilities to extend. How it extend is determined by the (L$\Pi \lambda$)
of the wavefunction. Thus, in the coordinate space, the OCTA\ is a source
where the INS may emerge and extend to the neighborhood surrounding the
OCTA. This fact implies that specific inherent nodal structure exists. The
details of the inherent nodal structure will not be concerned in this paper.
However, it is emphasized that for a wavefunction, if the OCTA is
accessible, all the shapes in the neighborhood of the OCTA are also
accessible, therefore this wavefunction is inherent nodeless in this domain.

Another shape with also a stronger geometric symmetry is a regular
centered-pentagons(C-PENTA, the particle 6 is assumed to be located at the
center of mass O). Let k' be the 5-fold axis. The C-PENTA is invariant to
(i) a rotation about k' by $\frac{2\pi }5$ together with a cyclic
permutation of the five particles of the pentagon , (ii) a rotation about k'
by $\pi $ together with a space inversion, (iii) a rotation about i' by $\pi 
$ together with $p_{14}p_{23}$ (here i' is the axis vertical to k' and
connecting O and particle 5). These invariances will lead to constraints
embodied by sets of homogeneous equations, and therefore the accessibility
of the C-PENTA can be identified as also given in Tab.2.

In addition to the OCTA, the C-PENTA is another source where the INS may
emerge and extend to its neighborhood; e.g., extend to the pentagon-pyramid
as shown in Fig.1b with h$\neq $0. There are also other sources. For
example, the one at the regular hexagons. However, among the 15 bonds, 12
can be optimized at an OCTA, 10 at a pentagon-pyramid, but only 6 at a
hexagon. Therefore in the neighborhood of the hexagon (and also other
regular shapes) the total potential energy is considerably higher. Since the
wavefunctions of the low-lying states are mainly distributed in the domain
with a relatively lower potential energy, we shall concentrate only in the
domains surrounding the OCTA\ and the C-PENTA.

When (L$\Pi \lambda $) =(0+\{6\}), (0+\{4,2\}), or (0+\{2,2,2\}), the
wavefunction can access both the OCTA and the C-PENTA (refer to Table 2).
These and only these wavefunctions are inherent-nodeless in the two most
important domains, and they should be the dominant components for the
low-lying states. All the other L=0 components must contain at least an INS
resulting in a great increase in energy. From Tab.1 it is clear that the
(0+\{6\}) component is not allowed, while the (0+\{4,2\}) component can be
contained in [S,T]=[1,0] and [0,1] states, and the (0+\{2,2,2\}) component
can be contained in [S,T]=[1,0], [3,0], [0,1], and [2,1] states. When
[S,T]=[1,0] , the $\lambda $ can be \{4,2\} or \{2,2,2\}, therefore two J$%
^\Pi =1^{+}$ partner-states with their spatial wavefunctions orthogonal to
each other exist, each of them is a specific mixture of \{4,2\} and
\{2,2,2\}. Similarly, two partner-states with [S,T]=[0,1] and J$^\Pi =0^{+}$
exist also. When [S,T]=[3,0] or [2,1], the $\lambda $ has only one choice,
therefore in each case only one state exists. Thus we can predict that there
are totally six low-lying states dominated by L=0 components without nodal
surfaces as listed in Tab.3, where the L,S, and $\lambda $ are only the
quantum numbers of the dominant component.

\hspace{1.0in}

\begin{tabular}{|cccccc|c|}
\hline
\multicolumn{1}{|c|}{S} & \multicolumn{1}{c|}{T} & \multicolumn{1}{c|}{J} & 
\multicolumn{1}{c|}{$\Pi $} & \multicolumn{1}{c|}{L} & $\lambda $ & E \\ 
\hline
\multicolumn{1}{|c|}{1} & \multicolumn{1}{c|}{0} & \multicolumn{1}{c|}{1} & 
\multicolumn{1}{c|}{+} & \multicolumn{1}{c|}{0} & \{4,2\} and \{2,2,2\} & 0
\\ \hline
\multicolumn{1}{|c|}{1} & \multicolumn{1}{c|}{0} & \multicolumn{1}{c|}{1} & 
\multicolumn{1}{c|}{+} & \multicolumn{1}{c|}{0} & \{4.2\} and \{2,2,2\} & 
5.65 \\ \hline
\multicolumn{1}{|c|}{3} & \multicolumn{1}{c|}{0} & \multicolumn{1}{c|}{3} & 
\multicolumn{1}{c|}{+} & \multicolumn{1}{c|}{0} & \{2,2,2\} & 2.19 \\ \hline
& 0 & 2 & + &  &  & 4.31 \\ \hline
\multicolumn{1}{|c|}{0} & \multicolumn{1}{c|}{1} & \multicolumn{1}{c|}{0} & 
\multicolumn{1}{c|}{+} & \multicolumn{1}{c|}{0} & \{4,2\} and \{2,2,2\} & 
3.56 \\ \hline
\multicolumn{1}{|c|}{0} & \multicolumn{1}{c|}{1} & \multicolumn{1}{c|}{0} & 
\multicolumn{1}{c|}{+} & \multicolumn{1}{c|}{0} & \{4,2\} and \{2,2,2\} & 
\\ \hline
\multicolumn{1}{|c|}{2} & \multicolumn{1}{c|}{1} & \multicolumn{1}{c|}{2} & 
\multicolumn{1}{c|}{+} & \multicolumn{1}{c|}{0} & \{2,2,2\} & 5.37 \\ \hline
\end{tabular}

Tab.3, Prediction of the quantum numbers of low-lying states (dominated by
L=0 components) of the 6-nucleon systems based on symmetry. The last column
is the energies (in MeV) of the states of $^6$Li taken from [11].

\hspace{1.0in}

It is expected that these low-lying states should be split by the nuclear
force. Owing to the interference of the \{4,2\} and \{2,2,2\} components,
there would be an larger energy gap lying between the two partner-states of
each pair. Ajzenberg-selove has made an analysis on $^6$Li based on
experimental data [11], the results are listed in Tab.3. Although our
analysis is based simply on symmetry, but the results of the two analyses
are close. For the T=0 states, there are two J$^\Pi =$ 1$^{+}$ states
([S,T]=[1,0]) in [11] with a split, they are just the expected partners. The
split is so large (5.65 MeV) that the lower one becomes the ground state
while the higher one becomes the highest state of this group. There is a T=0
state in [11] at 2.19 MeV with exactly the predicted quantum numbers J$^\Pi
=3^{+}$. Nonetheless, there is a T=0 state in [11] at 4.31 MeV with J$^\Pi $
= 2$^{+}$, which do not appear in our analysis. May be this state is
dominated by L=1 component, may be there is another origin to be clarified.

For the T=1 states, one of the expected partners with J$^\Pi =0^{+}$
([S,T]=[0,1]) was found in [11] at 3.56 MeV . However, the other partner (
it would be considerably higher) has not yet been identified in [11], this
is an open problem. Nonetheless, if this state exists, the structure of its
spatial wavefunction would be similar to the T=0 state at 5.65 MeV . The
third expected T=1 state was found in [11] at 5.37MeV with exactly the
predicted J$^\Pi =2^{+}$.

In summary we have explained the origin of the quantum numbers of the
low-lying states of 6-nucleon systems. The explanation is very different
from that based on the shell model [12,13]. For example, according to our
analysis, the J$^\Pi =3^{+}$ state at 2.19 MeV has S=3 and L=0. On the
contrary, in the shell model the four nucleons in the 1s orbit must have
their total spin zero and total isospin zero; therefore this state should
have S $\leq 1$ and L $\geq 2.$ However, it is noted that the 2$_1^{+}$
state (having S=0 and L=2) of the $^{12}C$ lies at 4.44 MeV [14]. Since the $%
^6$Li is considerably lighter and smaller than the $^{12}$C, the L=2 state
of $^6$Li should be much higher than 4.44 MeV due to having a much smaller
moment of inertia. Therefore the 3$^{+}$ state at 2.19 MeV is difficult to
be explained as a L $\geq $ 2 state. In particular, it is found that the
\{2,2,2\} component is important; however this component is suppressed by
the shell model. Thus, our analysis raises a challenge to the shell model in
the case that the number of nucleons is not large enough. Evidently, much
work should be done to clarify the physics underlying these systems.

It has been shown that sources of INS\ may exist in the quantum states.
Nonetheless, there are essentially inherent-nodeless components of
wavefunctions (each with a specific set of (L$\Pi \lambda $)). They are the
most important building blocks to constitute the low-lying states. The
identification of these particularly favorable components is a key to
understand the low-lying spectrum.

The idea of this paper can be generalized to investigate different kinds of
systems, thereby we can understand them in an unified way.

\hspace{1.0in}

ACKNOWLEDGEMENT: This work is supported by the NNSF of the PRC, and by a
fund from the National Educational Committee of the PRC.

\hspace{1.0in}

REFERENCES

1, B.S.Pudliner, V.R. Pandharipande, J.Carlson, and R.B.Wiringa, Phys. Rev.
Lett. 74, 4396, (1995)

2, B.S.Pudliner, V.R. Pandharipande, J.Carlson, S.C.Pieper, and R.B.Wiringa,
Phys. Rev. C56, 1720 (1997)

3, K.Varga, Y.Suzuki, Phys. Rev. C52, 2885, (1995)

4, A.Cs\'ot\'o, Phys. Rev. C49, 3035, (1994)

5, Y. Fujiwara and Y.C.Tang, Phys. Rev.C43, 96, 1991; Few-Body Systems 12,
21, (1992.)

6, C.G.Bao, Conference Handbook of XVth International Conference on Few-Body
Problems in Physics, edited by L.P.Kok, J.C.S.Bacelar, and A.E.L.Dieperink,
Gr\"oningen, p.496, 1997: nucl-th/9805001(preprint).

7, C.G.Bao, Few-Body Systems, 13, 41, (1992); Phys. Rev. A47, 1752 , (1993);
Phys. Rev. A50, 2182, (1994); Chinese Phys. Lett. 14, 20, (1997); Phys. Rev.
Lett., 79, 3475,(1997.)

8, W.Y.Ruan, and C.G.Bao, Few-Body Systems, 14, 25, (1993)

9, C.Itzykson and M.Nauenberg, Rev. Mod. Phys. 38, 95, (1966)

10, J.Q.Chen, ''Group Representation Theory for Physicists'', World
Scientific, Singapore ,1989

11, F.Ajzenberg-Selove, Nucl.Phys. A490, 1, 1(988)

12, M.G.Mayer and J.H.D.Jensen, ''Elementary Theory of Nuclear Shell
Structure'', Willey, New York, 1955

13, A. deShalit and I.Talmi, ''Nuclear Shell Theory'', Academic, New York,
1963

14, Y.Fujiwara, H.Horiuchi, K.Ikeda, M.Kamimura, K.Kato, Y.Suzuki, and
E.Uegaki, Prog. Theor. Phys., Supplement 68, 29, (1980)

\end{document}